\documentclass{epl}

\title{Bloch oscillations of strongly interacting Bose atoms}
\author{Andrey R. Kolovsky\inst{1,2}}
\institute{
\inst{1} Max-Planck-Institut f\"ur Physik komplexer Systeme, 01187 Dresden, Germany\\
\inst{2} Kirensky Institute of Physics, 660036 Krasnoyarsk, Russia}

\pacs{05.60.Gg}{Quantum transport}
\pacs{05.30.Jp}{Boson systems}
\pacs{03.65.-w}{Quantum mechanics}

\begin{document}
\maketitle

\begin{abstract}
We analyse Bloch oscillations (i.e., oscillations induced  by a static
force) of strongly interacting Bose atoms {\em beyond} the hard-core
bosons model. It is shown that residual interactions between the atoms
modulate Bloch oscillations, where the type of modulations depends on
the magnitude of the static force.
\end{abstract}


{\em 1.} Recently much attention has been paid to the transport
of cold atoms in optical lattices
\cite{Daha96,Mors01,Cata03,60,57,63,Scot04,Ott04,Pezz04,Fert05,Rigo05,70}.
One of these problems is Bloch oscillations (BO) of atoms
in a static, for example, gravitational field
(see review \cite{63} and references therein).
Besides their own interest, these studies pursue the aim
of realizing (and understanding) the ordinary and super-conductivity
of cold atoms \cite{Ott04,70}.

The present work is devoted to
BO of Bose atoms in deep anisotropic optical lattices, where
the system can be described by the 1D Bose-Hubbard model
\begin{equation}
\label{1}
\widehat{H}=\widehat{H}_0+Fd\sum_l l \hat{n}_l \;,
\end{equation}
\begin{equation}
\label{2}
\widehat{H}_0=
-\frac{J}{2}\sum_l \left( \hat{a}^\dag_{l+1}\hat{a}_l+h.c.\right)
  +\frac{W}{2}\sum_l \hat{n}_l(\hat{n_l}-1) \;.
\end{equation}
In Eqs.~(\ref{1}-\ref{2}) $\hat{a}^\dag_l$ and $\hat{a}_l$
are the bosonic creation and annihilation operators,
$\hat{n}_l=\hat{a}^\dag_l\hat{a}_l$ the number operator,
$J$ the hopping matrix element, $W$ the on-site interaction energy,
$d$ the lattice period, and $F$ the magnitude of static force.
For vanishing interactions the static force in (\ref{1})
induces coherent, perfectly periodic oscillations of the atoms
with the Bloch frequency $\omega_B=Fd/\hbar$. Then the main question
is of how non-zero interactions do affect these
oscillations.

For a weak to moderate interaction constant, $W\le J$, BO of Bose atoms
in 1D lattices were analysed in Ref.~\cite{60}.
It was found that atom-atom interactions destroy the coherence
in the system and BO typically decay within a few Bloch periods.
(An exception is the case of strong forcing, $Fd\gg J$, where BO of
interacting atoms are quasiperiodic \cite{57}.) One might naively
expect that stronger interactions would cause an even faster decay of BO.
This is, however, not the case because for $W\gg J$ the system
approaches the hard-core bosons (HC-bosons) limit,
\begin{equation}
\label{3}
\widehat{H}_{HC}=
-\frac{J}{2}\sum_l \left( \hat{a}^\dag_{l+1}\hat{a}_l+h.c.\right) \;,
\end{equation}
where the additional constraints $(\hat{a}_l)^2=(\hat{a}^\dag_l)^2=0$
prohibit double occupancy of a single well. The system (\ref{3}) is
completely integrable and its Hamiltonian can be diagonalised
by using the canonical transformation
$\hat{b}_k=(1/\sqrt{L})\sum_l \exp(i 2\pi k l/L)\hat{a}_l$, i.e., by
going from the Wannier to Bloch basis. Then it is easy to show
that BO of HC-bosons do not decay [see Eq.~(\ref{8}) below].

The above result refers to the formal limit $W\rightarrow\infty$.
In practice, however, one deals with a finite interaction constant,
which implies the presence of residual interactions between Bose
atoms. It is the primary aim of this work to study the effect of these
residual interactions on the Bloch dynamics of strongly interacting
Bose atoms.

\begin{figure}[t]
\center
\includegraphics[width=9cm]{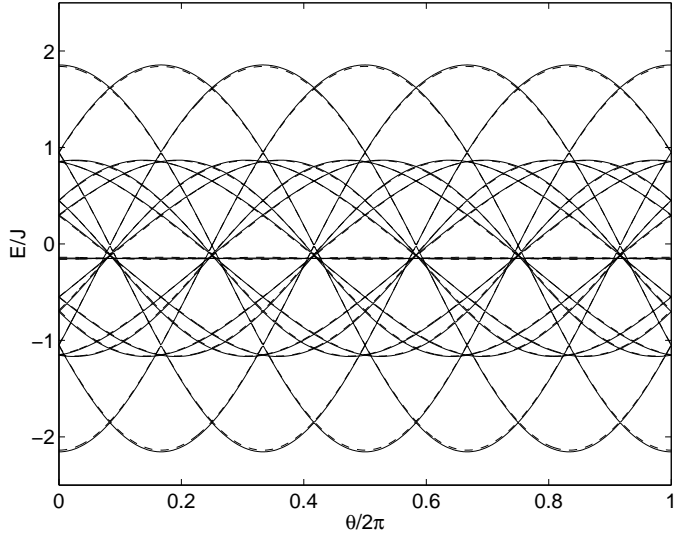}
\caption{Spectrum of the SC-bosons Hamiltonian (\ref{5}) parameterised by
the phase $\theta$ (solid lines) as compared to the low-energy spectrum 
of the Hamiltonian (\ref{2}) (dashed lines). The parameters are
$L=6$, $N=3$, and $W=10J$.}
\label{fig1}
\end{figure}

\bigskip
{\em 2.} We begin with the effective Hamiltonian for strongly interacting
Bose atoms\footnote{
The derivation of the effective Hamiltonian presented in this work is similar to
that for the effective Hamiltonian of composed bosons in the frame of the
negative-$u$ Fermi-Hubbard model (see, Ref.~\cite{Micn90}, for example).}
.
Assuming periodic boundary conditions
the Hilbert space ${\cal H}_{HC}$ of the system (\ref{3}) is spanned by
the Fock states $|{\bf n}\rangle=|n_1,\ldots,n_L\rangle$ with the occupation
numbers $n_l$ being ether zero or unity. To take the residual interactions
into account we extend ${\cal H}_{HC}$ by including the Fock states with $n_l=2$,
\begin{equation}
\label{4}
\widehat{H}^{(1+2)}=\left|
\begin{array}{cc}
\widehat{H}^{(2)}&\widehat{V}\\
\widehat{V}^\dag&\widehat{H}_{HC}
\end{array}
\right| \;.
\end{equation}
In the Hamiltonian (\ref{4}) the coupling matrix $\widehat{V}$ is 
proportional to $J$ and $\widehat{H}^{(2)}|{\bf n}\rangle=W|{\bf n}\rangle$.
Using ordinary perturbation theory in the parameter $J/W\ll 1$ and projecting the
eigenstates of (\ref{4}) back to the Hilbert space ${\cal H}_{HC}$,
we end up with the following effective Hamiltonian,
\begin{displaymath}
\widehat{H}_{SC}=
-\frac{J}{2}\sum_l \left( \hat{a}^\dag_{l+1}\hat{a}_l+h.c.\right)
\end{displaymath}
\begin{equation}
\label{5}
-\frac{J^2}{W}\sum_l\hat{n}_{l+1}\hat{n}_l
-\frac{J^2}{2W}\left(\sum_l \hat{a}^\dag_{l+1}\hat{n}_l\hat{a}_{l-1}+h.c.\right) \;.
\end{equation}
For finite $W$ the `soft-core bosons' (SC-bosons) Hamiltonian (\ref{5}) 
approximates the low-energy spectrum of Bose atoms much better than
the HC-bosons Hamiltonian (\ref{3}). As an example, Fig.~\ref{fig1} compares 
the low-energy spectrum of the system (\ref{2}) with the spectrum of
the SC-bosons Hamiltonian for $J/W=0.1$ where, to facilitate
the comparison, we parameterised the creation and annihilation operators
by the phase $\theta$: $\hat{a}^\dag_l\rightarrow\hat{a}^\dag_l\exp(-il\theta)$,
$\hat{a}_l\rightarrow\hat{a}_l\exp(il\theta)$. (The reason for this
parameterisation will become clear in a moment.) Note that some of the level
crossings in Fig.~\ref{fig1} are actually avoided crossings. We found,
however, that in the relevant region $J/W<0.1$ the gaps of
the avoided crossings can be safely neglected. This result also means
that the single-particle quasimomentum numbers $k_i$, $i=1,N$, are still
good quantum numbers to label the eigenstates of the Hamiltonian (\ref{5}).

\bigskip
{\em 3.} We proceed with the Bloch dynamics. Using a gauge transformation
the static forcing can be presented as a periodic driving
of the system with the Bloch frequency $\omega_B$,
\begin{eqnarray}
\label{5a}
\widehat{H}_{SC}(t)&=&\widehat{H}_{HC}(t)+\widehat{H}_{int}(t) \;,
\\
\nonumber
\widehat{H}_{HC}(t)&=&
-\frac{J}{2}\sum_l \left(e^{-i\omega_B t} \hat{a}^\dag_{l+1}\hat{a}_l+h.c.\right) \;,
\\
\nonumber
\widehat{H}_{int}(t)&=&-\frac{J^2}{W}\sum_l \hat{n}_{l+1}\hat{n}_l
-\frac{J^2}{2W}\sum_l\left(e^{-i2\omega_B t}\hat{a}^\dag_{l+1}\hat{n}_l\hat{a}_{l-1}+h.c.\right) \;.
\end{eqnarray}
Then the dynamics of the system is conveniently described by the evolution operator
over one Bloch period $\widehat{U}_{SC}(T_B)$, where
\begin{equation}
\label{6}
\widehat{U}_{SC}(t)=\widehat{\exp}\left[-\frac{i}{\hbar}
\int_0^t \widehat{H}_{SC}(t) dt \right]
\end{equation}
and the hat over the exponent sign denotes the time odering. For $W=\infty$
(the HC-bosons limit) the explicite form of the evolution operator is given by
\begin{equation}
\label{7}
\widehat{U}_{HC}(t)=\widehat{T}^\dag \widehat{D}(t) \widehat{T} \;,
\end{equation}
where the unitary operator $\widehat{T}$ represents the transformation from
the Wannier basis $|{\bf n}\rangle=|n_1,\ldots,n_L\rangle$
to the Bloch basis $|{\bf k}\rangle=|k_1,\ldots,k_L\rangle$ \footnote{
The elements of $T$ are given by the determinant of
the $N\times N$ matrix $A$ with $A_{i,j}=L^{-1/2}\exp(ik_i l_j/L)$,
where $k_i$ and $l_j$ are the occupied orbitals.}
and the matrix of the operator $\widehat{D}(t)$ is diagonal in the Bloch basis,
\begin{equation}
\label{8}
\langle {\bf k}|D(t)|{\bf k}\rangle =
\exp\left[-i\frac{J}{\hbar}\int_0^t\sum_{i=1}^N
\cos\left(\frac{2\pi k_i}{L}-\omega_B t\right) dt\right] \;.
\end{equation}
From (\ref{7}-\ref{8}) it follows immediately
that $\widehat{U}_{HC}(T_B)=\widehat{I}$ and, hence, BO of HC-bosons are
perfectly periodic. In contrast, for SC-bosons
the Bloch evolution operator $\widehat{U}_{SC}(T_B)$ differs from the identity
operator and, thus, BO are generally not periodic.

\begin{figure}[t]
\center
\includegraphics[width=9cm]{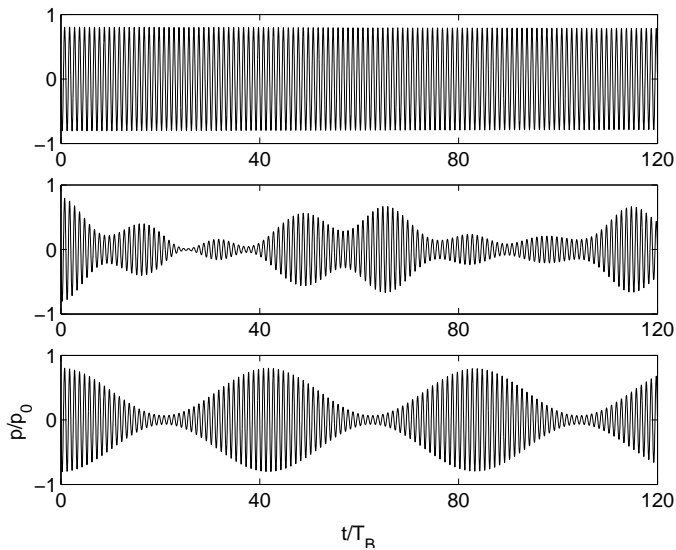}
\caption{Normalised mean momentum ($p(t)\rightarrow p(t)/p_0$, $p_0=NJMd/\hbar$)
of strongly interacting Bose atoms for
weak ($Fd/J=0.1$, upper panel), moderate ($Fd/J=1$ middle panel),
and strong ($Fd/J=4$, lower panel) forcing. The other parameters are
$L=8$, $N=3$, and $W=10J$.}
\label{fig2}
\end{figure}

\bigskip
Addressing the dynamics of interacting Bose atoms one has to distinguish
between strong, $Fd\gg J$, and weak, $Fd\ll J$, forces. We discuss a weak force first.
In this case the phase $\theta=\omega_B t$ in Eq.~(\ref{5a})
can be considered as a slowly varying parameter and, hence, the system adiabatically follows
the instantaneous levels shown in Fig.~\ref{fig1}. Then the matrix of the evolution
operator $\widehat{U}_{SC}(T_B)$ is {\em diagonal} in the eigen-basis of the Hamiltonian
$\widehat{H}_{SC}$. For $J/W\ll 1$ this basis practically coincides with the eigen-basis
of $\widehat{H}_{HC}$, i.e., with the quasimomentum Fock states $|{\bf k}\rangle$.
Thus
\begin{equation}
\label{9}
\langle {\bf k}|\widehat{U}_{SC}(T_B)|{\bf k}\rangle=
\exp[-i\Theta({\bf k})] \;,
\end{equation}
where $\Theta({\bf k})=\frac{1}{\hbar}\int_0^{T_B} E_{\bf k}(t) dt$
are the phases acquired  by the states during the time evolution.
Since the energy differences between the levels of the HC-bosons
and SC-bosons Hamiltonians are essentially given by the diagonal elements
of $\widehat{H}_{int}(t)$ in the Bloch basis, we finally have
\begin{equation}
\label{11}
\Theta({\bf k})=-\frac{2\pi J^2}{W Fd}
\langle {\bf k}|\sum_l \hat{n}_{l+1}\hat{n}_l |{\bf k}\rangle \;.
\end{equation}
Assuming the experimentally relevant situation where the system is initially
in its ground state, the result (\ref{9}-\ref{11}) means that BO of SC-bosons
are periodic in the limit of small $F$. This regime of the system's dynamics
is illustrated in the upper panel of Fig.~\ref{fig2}, showing the
mean atomic momentum of $N=3$ atoms in $L=8$ wells (periodic boundary
conditions) for $J/W=0.1$ and $Fd/J=0.1$.

\begin{figure}[t]
\center
\includegraphics[width=9cm]{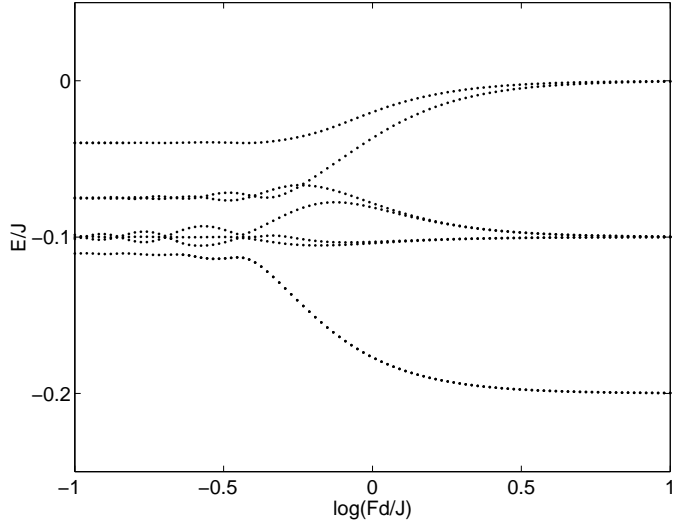}
\caption{Quasienergy spectrum of the system for different magnitudes of 
the static force. The system parameters are $L=8$, $N=3$, and $W=10J$.}
\label{fig3}
\end{figure}

\bigskip
When the static force is increased, the adiabatic approximation breaks down
and the Bloch dynamics of the atoms becomes rather complicated. An example
is given in the middle panel in Fig.~\ref{fig2}, where the static force
$Fd/J=1$. It is seen that the mean momentum $p(t)$ undergoes a complex
quasiperiodic process, involving many different frequencies. These
frequencies are obviously given by the quasienergies $E$,
which are defined by the eigen-phases $\Theta$ of the evolution operator,
\begin{equation}
\label{11a}
\widehat{U}_{SC}(T_B)|\Phi\rangle=\exp(-i\Theta)|\Phi\rangle \;,
\end{equation}
through the relation $\Theta=E T_B/\hbar=2\pi E/\hbar\omega_B$.
For the considered system of $N=3$ atoms in $L=8$ wells the quasienergies
$E$ are plotted in Fig.~\ref{fig3} as the functions of $F$, the strength of
the static force. The transition from the adiabatic to the non-adiabatic regime,
reflected in the removed degeneracy between the quasienergy levels,
is clearly seen\footnote{
Still, since the global quasimomentum is conserved, each of the
quasienergy levels is $L$-fold degenerate.}.
%
It is also an appropriate place here to mention the particle-hole symmetry
of the system. In terms of the evolution operator it means that the
quasienergy spectum of $N=3$ atoms in $L=8$ wells coincides (up to an irrelevant
global shift) with the spectrum of $N=5$ atoms in $L=8$ wells.

\bigskip
The further increase of $F$ simplifies the Bloch dynamics again, as can already
be concluded from the structure of the quasienergy spectrum for $Fd\gg J$ (see Fig.~\ref{fig3}).
Let us show that in this region BO of SC-bosons are periodically modulated
with the period
\begin{equation}
\label{12}
T_W=\frac{2\pi \hbar W}{J^2}=\frac{W}{J}\frac{Fd}{J} T_B
\end{equation}
(see lower panel in Fig.~\ref{fig2}).  Indeed, for strong forcing the Bloch
period $T_B$ is the shortest time scale in the system and, hence, one can
separate the slow (modulation) and fast (BO) dynamics. Introducing the
`slow' wave-function $|\widetilde{\Psi}(t)\rangle$,
$|\Psi(t)\rangle=\widehat{U}_{HC}(t)|\widetilde{\Psi}(t)\rangle$,
the Schr\"odinger equation reads
\begin{equation}
\label{13}
i\hbar\frac{\partial |\widetilde{\Psi}\rangle}{\partial t}
=\widehat{H}_{slow}|\widetilde{\Psi}\rangle \;,
\end{equation}
%
\begin{equation}
\label{14}
\widehat{H}_{slow}=\frac{1}{T_B}\int_0^{T_B}\widehat{U}_{HC}^\dag(t)
\widehat{H}_{int}(t)\widehat{U}_{HC}(t) dt \;,
\end{equation}
where $\widehat{U}_{HC}(t)$ is given in Eqs.~(\ref{7}-\ref{8}). Noting that for $Fd\gg J$
the operator $\widehat{D}(t)\approx\widehat{I}$, we conclude that the
matrix of the evolution operator $\widehat{U}_{SC}(T_B)$ is {\em diagonal} in the Wannier
basis,
\begin{equation}
\label{15}
\langle {\bf n}|\widehat{U}_{SC}(T_B)|{\bf n}\rangle=
\exp[-i\Theta({\bf n})] \;,
\end{equation}
%
\begin{equation}
\label{16}
\Theta({\bf n})=-\frac{2\pi J^2}{W Fd}
\langle {\bf n}|\sum_l \hat{n}_{l+1}\hat{n}_l |{\bf n}\rangle \;.
\end{equation}
Furthermore, because the quantities
$\langle{\bf n}|\sum_l \hat{n}_{l+1}\hat{n}_l |{\bf n}\rangle$
are integers, the relation (\ref{15})
ensures perfectly periodic modulations of the BO,
\begin{equation}
\label{17}
p(t)=[A+B\cos(2\pi t/T_W)]\sin(\omega_B t) \;.
\end{equation}
The constants $A$ and $B$ in Eq.~(\ref{17}) depend on the filling factor
${\bar n}=N/L$ and one gets the most pronounced modulations for
${\bar n}\sim 1/2$ \footnote{
The functional dependence (\ref{17}) also holds
for BO of two-component Fermi gas (repulsive interactions) in the limit
of strong forcing $Fd\gg J$ \cite{72}.}.
It is also worth of mentioning that the above analysis implicitely
assumes $Fd<W$. Indeed, for $Fd\approx W$ the static forcing
{\em resonantly} couples the Fock states belonging to ${\cal H}_{HC}$
to the states with $n_l=2$ and our approach (based on ordinary
perturbation theory) is not valid. For a particular case of
the Mott-insulator initial state (${\bar n}=1$) the dynamical
responce of the system to the resonant static force was analysed
in Refs.~\cite{Brau02,Sach02,64}.

\bigskip
{\em 4.} In conclusion, we have studied the effect of residual atom-atom
interactions on the Bloch dynamics of strongly interacting Bose atoms.
The characteristic energy of these interactions is given by the
ratio of the squared hopping matrix element $J$ and the microscopic
interaction constant $W$, and is usually neglected when one discusses
the equilibrium properties of the system (i.e., one uses the hard-core
bosons model \cite{Pare04,Rigo04b}). The effect of residual interactions, 
however, accumulates in time when it concerns the dynamics.
In particular, it has been shown that they may modulate BO
of Bose atoms. We have provided a complete description of
the dynamical regimes of the system in terms of the Floquet-Bloch 
evolution operator (evolution operator over one Bloch period
$T_B=2\pi\hbar/Fd$). The matrix of this operator is found to be
diagonal in the Bloch basis in the case of a weak forcing, $Fd\ll J$,
diagonal in the Wannier basis in the opposite case
$Fd\gg J$, and non-diagonal in either of basis for a moderate forcing,
$Fd\sim J$. As the result, BO of the strongly
interacting Bose atoms are periodic for a weak static force,
periodically modulated with the period $T_W=2\pi\hbar W/J^2$
for a strong static force, and involve multiple time scales for a moderate
static force.

It is interesting to estimate the modulation period $T_W$ in the typical laboratory
experiment on BO of Bose atoms in the gravitational field. Taking, as an example, rubidium
atoms in a 3D optical lattice with $d=0.405\mu$m ($E_R/2\pi\hbar=3558$Hz), $V_z=6E_R$,
and $V_\perp=30E_R$, we have $J^2/W=0.1^2/1.3\approx0.008E_R$ and the Stark
energy $Fd=0.24E_R$. Thus the modulation period $T_W=30T_B=35$ms.
This time is easily accessible in the present day experiments.
It should be noted, however, that the results of the present work refer
to a finite lattice with periodic boundary conditions, while in practice
one typically uses a harmonic confinement of the system.
The analysis of the Bloch dynamics of Bose atoms for these different boundary
conditions will be the subject of a separate paper.



\begin{thebibliography}{10}


\bibitem{Daha96}
\Name{M.~Ben Dahan, E.~Peik, J.~Reichel, Y.Castin \and C.~Salomon}
\REVIEW{Phys. Rev. Lett.}{76}{1996}{4508}.

\bibitem{Mors01}
\Name{O.~Morsch, J.~H.~M\"uller, M.~Cristani, D.~Ciampini \and E.~Arimondo}
\REVIEW{Phys. Rev. Lett.}{87}{2001}{140402}.

\bibitem{Cata03}
\Name{F.~S.~Cataliotti, L.~Fallani, F.~Ferlaino, C.~Fort, P.~Maddaloni \and M.~Inguscio}
\REVIEW{New J. of Phys.}{5}{2003}{71.1}.

\bibitem{60}
\Name{A.~R.~Kolovsky \and A.~Buchleitner}
\REVIEW{Phys. Rev. E}{68}{2003}{056213};
%
\REVIEW{Phys. Rev. Lett.}{91}{2003}{253002}.

\bibitem{57}
\Name{A.~R.~Kolovsky}
\REVIEW{Phys. Rev. Lett.}{90}{2003}{213002}.

\bibitem{63}
\Name{A.~R.~Kolovsky \and H.~J.~Korsch}
\REVIEW{Int. J. of Mod. Phys.}{18}{2004}{1235}.

\bibitem{Scot04}
\Name{R.~G.~Scott, A.~M.~Martin, S.~Bujkiewicz, T.~M.~Fromhold,
N.~Malossi, O.~Morsch, M.~Cristiani \and E.~Arimondo}
\REVIEW{Phys. Rev. A}{69}{2004}{033605}.

\bibitem{Ott04}
\Name{H.~Ott, E.~de~Mirandes, F.~Ferlaino, G.~Roati, G.~Modugno \and M. Inguscio}
\REVIEW{Phys. Rev. Lett.}{92}{2004}{160601}.

\bibitem{Pezz04}
\Name{L.~Pezze, L.~Pitaevskii, A.~Smerzi \and S.~Stringari,
G.~Modugno, E.~de~Mirandes, F.~Ferlaino, H.~Ott, G.~Roati \and M. Inguscio}
\REVIEW{Phys. Rev. Lett.}{93}{2004}{120401}.

\bibitem{Fert05}
\Name{C.~D.~Fertig, K.~M.~O'Hara, J.~H.~Huckans, S.~L.~Rolston, W.~D.~Phillips \and J.~V.~Porto}
\REVIEW{Phys. Rev. Lett.}{94}{2005}{120403}.

\bibitem{Rigo05}
\Name{M.~Rigol, V.~Rousseau, R.~T.~Scalettar \and R.~R.~P.~Singh}
\REVIEW{Phys. Rev. Lett.}{95}{2005}{110402}.

\bibitem{70}
\Name{A.~V.~Ponomarev, J.~Mandro\~nero, A.~R.~Kolovsky \and A.~Buchleitner}
{\em Atomic current across an optical lattice}, to be published in
\REVIEW{Phys. Rev. Lett.}{96}{2006}.

\bibitem{Micn90}
\Name{R.~Micnas, J.~Ranninger \and S.~Robaszkiewicz}
\REVIEW{Rev. of Mod. Phys.}{62}{1990}{113}.

\bibitem{72}
\Name{A.~V.~Ponomarev, A.~R.~Kolovsky \and A.~Buchleitner}
unpublished.

\bibitem{Brau02}
\Name{K.~Braun-Munzinger, J.~A.~Dunningham \and K.~Burnett},
cond-mat/0211701 preprint, 2004.

\bibitem{Sach02}
\Name{Subir Sachdev, K.~Sengupta \and S.~M.~Girvin} 
\REVIEW{Phys. Rev. B}{66}{2002}{075128}.

\bibitem{64}
\Name{A.~R.~Kolovsky}
\REVIEW{Phys. Rev. A}{70}{2004}{015604}.

\bibitem{Pare04}
\Name{B.~Paredes, A.~Widera, V.~Murg, O.~Mandel, S.~F\"olling, I.~Cirac, G.~V.~Shlyapnikov,
T.~W.~H\"ansch \and I.~Bloch}
\REVIEW{Nature}{429}{2004}{277}.

\bibitem{Rigo04b}
\Name{M.~Rigol \and A.~Muramatsu}
\REVIEW{Phys. Rev. A}{70}{2004}{031603(R)}.

\end{thebibliography}
\end{document}